\documentclass[
prd,
superscriptaddress,
10 pt,
preprintnumbers,
notitlepage,
secnumarabic,
nobibnotes,
nofootinbib,
showpacs
]{revtex4-1}

\usepackage[T1]{fontenc}
\usepackage{bm}
\usepackage[pdftex]{color,graphicx}
\usepackage[makeroom]{cancel}
\usepackage{amssymb}
\usepackage{amsmath}
\usepackage{tabularx}
\usepackage{dsfont}
\usepackage{hyperref}
\usepackage{manfnt}
\usepackage{ulem}
\usepackage[stable]{footmisc}
\usepackage{tikz}
\usetikzlibrary{shapes}
\usetikzlibrary{arrows}
\tikzset{l/.style={draw=black, line width=1pt}}
\tikzset{s/.style={auto, outer sep=-1}}
\tikzset{t/.style={auto, outer sep=2, pos=-0.1}}
\tikzset{b/.style={auto, outer sep=2, pos=1.1}}
\tikzset{z/.style={auto, outer sep=-2}}
\tikzset{y/.style={auto, outer sep=1}}
\tikzset{v/.style={auto, outer sep=-2}}
\tikzset{h/.style={auto, outer sep=1}}
\tikzset{a/.style={pos=1.1}}
\tikzset{l1/.style={draw=black, line width=0.1pt}}
\tikzset{l2/.style={draw=black, line width=0.2pt}}
\tikzset{l4/.style={draw=black, line width=0.4pt}}
\tikzset{l6/.style={draw=black, line width=0.6pt}}

\usepackage{perpage}
\MakePerPage{footnote}

\newcommand{\bra}[1]{\ensuremath{\left\langle#1\right|}}
\newcommand{\ket}[1]{\ensuremath{\left|#1\right\rangle}}
\newcommand{\bracket}[2]{\ensuremath{\left\langle #1 \middle| #2 \right\rangle}}

\newcommand{\vket}[1]{\ensuremath{\Biggl|\vcenter{\hbox{#1}}\hspace{-5pt}\Biggr\rangle}}

\newcommand\atopp[2]{\genfrac{}{}{0pt}{}{#1}{#2}}
\DeclareMathAlphabet\mathbfcal{OMS}{cmsy}{b}{n}

\begin{document}
\title{Quantum reduced loop gravity: extension to gauge vector field}

\author{Jakub Bilski}
\email{jakubbilski14@fudan.edu.cn}
\affiliation{Center for Field Theory and Particle Physics \& Department of Physics, Fudan University, 200433 Shanghai, China}

\author{Emanuele Alesci}
\email{ealesci@sissa.it}
\affiliation{SISSA, Via Bonomea 265, 34126 Trieste and INFN Sez. Trieste, EU }

\author{Francesco Cianfrani} 
\email{francesco.cianfrani@ift.uni.wroc.pl}
\affiliation{Instytut Fizyki Teoretycznej, Uniwersytet Wroc\l{}awski, pl. Maksa Borna 9, 50-204 Wroc\l{}aw, EU}

\author{Pietro Don\`a} 
\email{pietro.dona@cpt.univ-mrs.fr}
\affiliation{Centre de Physique Theorique
Campus of Luminy, Case 907, F-13288 Marseille cedex 9, EU}

\author{Antonino Marcian\`o} 
\email{marciano@fudan.edu.cn}
\affiliation{Center for Field Theory and Particle Physics \& Department of Physics, Fudan University, 200433 Shanghai, China}

\begin{abstract}
\noindent
Within the framework of Quantum Reduced Loop Gravity we quantize the Hamiltonian for a gauge vector field. The regularization can be performed using tools analogous to the ones adopted in full Loop Quantum Gravity, while the matrix elements of the resulting operator between basis states are analytic coefficients. This analysis is the first step towards deriving the full quantum gravity corrections to the vector field semiclassical dynamics. 
\end{abstract}
\maketitle

\section{Introduction}
\noindent 
Quantum reduced loop gravity (QRLG) focuses on the quantization of gravitational syetems that are described by metrics with spatial part and dreibein gauge-fixed to a diagonal form. It was first introduced in \cite{Alesci:2012md,Alesci:2013lea} and then developed in \cite{Alesci:2013xd}-\cite{Bilski:2015dra} (see \cite{Alesci:2016gub} for a review). Hitherto the theory has been successfully implemented for the Bianchi I model. QRLG is derived from Loop Quantum Gravity (LQG) \cite{Ashtekar:2004eh,Rovelli:2004tv,Thiemann:2007zz} by imposing weakly gauge-fixing conditions to the states of the kinematical Hilbert space. Therefore it stands as the implementation of LQG to the cosmological sector. At the same time it differs from Loop Quantum Cosmology (LQC) \cite{Bojowald:2011zzb,Ashtekar:2011ni,Banerjee:2011qu}, in which the quantization of the gravitational system is performed in the minisuperspace, {\it i.e.} once symmetry-reduction of the phase space has taken place at a classical level. The semiclassical limit of QRLG reproduces the effective Hamiltonian of LQC \cite{Alesci:2014uha} with $\mu_0$ regularization, while the effective dynamics with improved regularization can be inferred via a statistical average over an ensemble of classically equivalent states \cite{Alesci:2016rmn}.

QRLG provides a novel derivation of earlier LQC results, in particular for what concerns the realization of the bouncing scenario. This picture still needs to be completed thou with the introduction of matter field. Since QRLG provides a true graph structure underlying the continuous Universe description, matter fields must be quantized via the tools of the loop quantization \cite{Thiemann:1997rq,Thiemann:1997rt}, {\it i.e.} in the full theory. Hence, QRLG provides an arena in which the implications of loop quantization for matter fields can be tested. The first of these analyses devoted to the study of matter in LQG traces back to \cite{Bilski:2015dra}, where the QRLG framework was extended in order to include a scalar field. We focus on the implementation of gauge vector fields in this framework. The interest for such a case is not only academic, given the potential role of vector fields in early cosmology \cite{Ford:1989me, Parker:1993ya, Golovnev:2008cf, Maleknejad:2011sq, Alexander:2011hz, Adshead:2012kp, Alexander:2014uza}, which could open a new window on quantum gravity phenomenology. Furthermore, the inclusion of gauge fields in LQG might also open new pathways toward a unified comprehension of forces, with peculiar phenomenological consequences --- see {\it e.g.} preliminary investigations in Ref.~\cite{Alexander:2011jf, Alexander:2012ge}.  

We first introduce vector fields in QRLG and define the operator corresponding to its contribution to the scalar constraint. We then quantize the field by adapting the LQG procedure given in \cite{Thiemann:1997rq,Thiemann:1997rt} (and recently reviewed in \cite{Liegener:2016mgc}) to the reduced model, similarly to the scalar field case discussed in \cite{Bilski:2015dra}. Basic quantum variables for the matter field are gauge group holonomies and fluxes, just like in lattice gauge theories \cite{Creutz}, the lattice being here provided by the cubic graph at which gravitational holonomies are based. The scalar constraint is quantized via a regularization of the classical expression, which provides a constraint for the gravitational contributions written entirely in terms of the holonomies and the fluxes proper of QRLG. Such a formulation is technically achievable and all the relevant computations can be performed analytically, thanks to the expressions of the matrix elements of the volume operator. The outcome of our analysis is the computation of the matrix elements (between the basis elements of QRLG) of the scalar constraint part that involves vector fields, and its expectation value on semiclassical states, which is the starting point for future applications.

In section \ref{II} we introduce QRLG and define the kinematical Hilbert space. We focus our attention on states based on graphs having six-valent nodes \cite{Alesci:2015nja}, which allow us to construct a cubulation of the whole spatial manifold. In section \ref{III}, the classical and quantum formulation for a vector field is given. On a classical level, we write the contributions of the vector field to the scalar, vector and Gauss constraints. The regularization of the field contribution to the scalar constraint is performed in section \ref{IV}. We just adapt to our case the regularization performed in \cite{Thiemann:1997rt}, {\it i.e.} we replace the triangulation with the cubulation of the spatial manifold and SU$(2)$ group elements of LQG with the corresponding U$(1)$ group elements in QRLG. Once the geometric variables have been cast in terms of fluxes and holonomies and the phase space coordinates of the vector field in terms of gauge group holonomies and fluxes, the quantization is straightforwardly performed in section \ref{V}. The resulting operator is discussed in the large $j$-limit. We find that when proper semiclassical states are constructed for the vector field, the expectation value of the vector field contribution to the scalar constraint reproduces the correspondent classical expression. This result provides a first check on the consistency of the adopted framework. 

Through all this paper we pick the conventions on metric signature $(-,+,+,+)$, on the gravitational coupling constant $\kappa=16\pi G$ and on the speed of light $c=1$. The metric tensor is defined as $g_{\mu\nu}=e^I_{\mu}e^J_{\nu}\eta_{IJ}$, where $e^I_{\mu}$ are vierbein fields and $\eta_{IJ}$ the flat metric. Dreibeins are denoted as $e^i_a$, where lowercase latin indexes $a,b,..=1,2,3$ label coordinate on each Cauchy hypersurface constructed by ADM decomposition \cite{Arnowitt:1962hi}, while $i,j,..=1,2,3$ are su$(2)$ internal indexes.

\section{Quantum reduced loop gravity}\label{II}
\noindent
The canonical variables of LQG are holonomies of Ashtekar-Barbero connections \cite{Ashtekar:1986yd}, smeared along some curve $\gamma$,
$h_{\gamma}\!:=\mathcal{P}\exp\!\left(\int_{\gamma}A^j_a(\gamma(s))\tau^j\dot{\gamma}^a(s)\right)$
 and fluxes of densitized triads across some surface $S$,
$E(S)\!:=\int_S n_j\epsilon_{abc}E^a_jdx^b\wedge dx^c$.
The kinematical Hilbert space of the theory is defined as the direct sum of the space of cylindrical functions of connections $\Psi_{\Gamma, f}(A):=\left<A\middle|\Gamma,f\right>:=f\big( h_{l^1}(A), h_{l^2}(A), ... , h_{l^L}(A) \big)$ along each graph $\Gamma$, with a continuous function
$f:{\rm SU}(2)^L\longrightarrow\mathds{C}$. It can be represented by the formula:
\begin{equation}
\mathcal{H}_{kin}^{(gr)}:=\bigoplus_{\Gamma}\mathcal{H}_{\Gamma}^{(gr)}\!=L_2\big(\mathcal{A},d\mu_{AL}\big),
\end{equation}
where $\mathcal{A}$ denotes the space of connections and $d\mu_{AL}$ is the Ashtekar-Lewandowski measure \cite{Ashtekar:1994mh}.

The basis states, called spin network states, are given by the expression:
\begin{equation}
\Psi_{\Gamma,j_l,i_v}(h)=\left<h\middle|\{\Gamma,{j_l},{i_v}\}\right>=\prod_{v\in\Gamma}i_v\cdot\prod_lD^{j_l}(h_l),
\end{equation}
where the triple $\{\Gamma,{j_l},{i_v}\}$ is a graph $\Gamma$, spin $j$ of the holonomy along each link $l$ and an intertwiner  $i_v$ implementing SU$(2)$ invariance at each node $v$. The product $\prod_l$ extends over all the links $l$ emanating from the node $v$ and $D^{j_l}(h_l)$ are the Wigner matrices, while the $\cdot$ denotes contraction of the SU$(2)$ indexes.

QRLG selects a finite amount of degrees of freedom in LQG and it can be applied to all gravitational models with diagonal metric tensor and triads, hence one can define the line element as
\begin{equation}
dl^2=a_1^2dx^2+a_2^2dy^2+a_3^2dz^2,
\end{equation}
where the three scale factors are functions of time and of all spatial coordinates. 
The graph $\Gamma$ now contains the set of links $l^i$, each being placed along a fiducial direction $1,2$ or $3$ . The canonical variables, which are un-smeared versions of holonomies and fluxes are:
\begin{equation}\label{diagonalmomentum}
E^i_a=p^i\delta^i_a,\quad A^i_a=c_i\delta^i_a
\end{equation}
(indexes are not summed in this expression), where $|p^i|=\frac{a_1a_2a_3}{a_i}$ and the reduced connections $c_i$ are proportional to the time derivatives of scale factors. This implies an SU$(2)$ gauge-fixing condition in the internal space, which is realized by the projection of SU(2) group elements, which are based at links $l^i$, onto U$(1)$ group representations. The U$(1)$ group elements are obtained by stabilizing the SU(2) group along the internal directions $\vec{u}_l=\vec{u}_i$, with 
\begin{equation}
\vec{u}_1=(1,0,0)\quad \vec{u}_2=(0,1,0) \quad \vec{u}_3=(0,0,1). 
\end{equation}
The kinematical Hilbert for QRLG reads:
\begin{equation}
^{R\!}\mathcal{H}_{kin}^{(gr)}:=\bigoplus_{\Gamma}\ \!\!^{R\!}\mathcal{H}_{\Gamma}^{(gr)}.
\end{equation}
with $\Gamma$ being a cuboidal graph. The basis states in the reduced Hilbert space $^{R\!}\mathcal{H}_{\Gamma}^{(gr)}$ are obtained by projecting SU(2) Wigner matrices on the state $|m_l,\vec{u}_l\rangle$, where $m_l$ is the magnetic number attached to the link $l$, which can take only maximum or minimum value ($m_l=\pm j_l$) for the angular momentum component $J_l=\vec{J}\cdot \vec{u}_l$:
\begin{equation}
^{l}\!D^{j_l}_{m_lm_l}(h_l)=\left<m_l,\vec{u}_l\middle|D^{j_l}(h_l)\middle|m_l,\vec{u}_l\right>,\ \ h_l\in\text{SU(2)}.
\end{equation}
Then the basic states, called reduced spin network states, are defined by the following expression:
\begin{equation}
^{R\!}\Psi_{\Gamma,m_l,i_v}(h)=\left<h\middle|\{\Gamma,m_l,i_v\}\right>
=\prod_{v\in\Gamma}\left<j_l,i_v\middle|m_l,\vec{u}_l\right>\cdot\prod_l\ \!\!^{l}\!D^{j_l}_{m_lm_l}(h_l), \ \ m_l=\pm j_l,
\end{equation}
where $\left<j_l,i_v\middle|m_l,\vec{u}_l\right>$ are reduced intertwiners of U$(1)$ group. The reduced spin network states are not orthogonal with respect to the intertwiners $i_v$, since the scalar product reads:
\begin{equation}\label{scalar_product}
\bracket{\Gamma,m_l,i_v}{\Gamma',m_l',i_v'}
=\delta_{\Gamma,\Gamma'}\prod_{v\in\Gamma}\prod_{l\in\Gamma}\delta_{m_l,m_l'}
\bracket{m_l,\vec{u}_l}{j_l,i_v}\bracket{j_l,i_v'}{m_l,\vec{u}_l}.
\end{equation}
It is convenient to normalize those states: in this way the reduced intertwiners being just phases drop out from the kinematical reduced Hilbert space.
\begin{equation}
\left|\Gamma,m_l,i_v\right>_N=\ket{\Gamma;m_l}_{\!R}
\end{equation}
where $_N$ stands for normalized. In this way the Hilbert space $^{R\!}\mathcal{H}_{\Gamma}^{(gr)}=\otimes_{l_i\in \Gamma} \mathcal{H}_{l_i}$ with $\mathcal{H}_{l_i}$ being the $U(1)_i$ Hilbert space associated to each link $l_i$ in direction $i$.
 The reduced intertwiners will then just appear when we project the $SU(2)$ operators.

The graphical way of constructing the elements of $\!\!^{R\!}\mathcal{H}_{\Gamma}^{(gr)}$ out of those of the full theory is to replace SU(2) basis elements with the following objects
\begin{equation}
\raisebox{-1.5ex}{
\begin{tikzpicture}
\draw[l] (-1.2,0) --  node[v] {$j_l$} (-0.7,0);
\draw[l] (-0.7,-0.1) -- (-0.7,0.1);
\draw[l] (-0.5,-0.1) -- (-0.5,0.1);
\draw[l] (-0.3,0) -- (-0.5,0);
\draw[l] (0,0) circle(3mm); \node at (0.02,0) {$h_l$};
\draw[l] (0.3,0) -- (0.5,0);
\draw[l] (0.5,-0.1) -- (0.5,0.1);
\draw[l] (0.7,-0.1) -- (0.7,0.1);
\draw[l] (0.7,0) --  node[v] {$j_l$} (1.2,0);
\end{tikzpicture}
}
=
\left<j_l,m\middle|m'',\vec{u}_l\right>\left<m'',\vec{u}_l\middle|D^{j_l}(h_l)\middle|m'',\vec{u}_l\right>\left<m'',\vec{u}_l\middle|j_l,m'\right>,\quad m''=\pm j_l.
\end{equation}

Finally, the canonical reduced variables are $^{R\!}h_{l^i}$ and $^{R\!}E(S)$, which are constructed by smearing along links of reduced, cuboidal graph $\Gamma$ and across surfaces $S$ perpendicular to these links, respectively. Since on the quantum level we use only reduced variables, we neglect the left uppercase symbol $^{R}$ in the next sections. The scalar constraint operator, neglecting the scalar curvature term, is obtained from that of LQG by considering only the euclidean part and replacing LQG operators with reduced ones. Its action on three-valent and six-valent nodes has been analyzed in \cite{Alesci:2013xd} and \cite{Alesci:2015nja}, respectively.

It is worth nothing that the nodes of the cubiodal graph $\Gamma$ are always six-valent and the graphical representation of a subsystem containing one node with attached links is given by the figure:
\begin{equation}\label{state_gravity}
\ket{\Gamma;j_l}_{\!R}
=\!\vket{
\begin{tikzpicture}
\draw[l] (-1.7,0) --  node[h] {$j_{x,y-1,z}^{(2)}$} (-0.2,0);
\draw[l] (-1.7,0) -- (-2.3,0);
\draw[l] (-0.2,-0.1) -- (-0.2,0.1);
\draw[l] (-2.45,0) circle(1.5mm); \node at (-2.4,0.4) {$h_{^{x\!,y\texttt{-\!}1\!,z}}^{\!(2)}$};
\draw[l] (-2.6,0) -- (-3.2,0);
\draw[l] (-4.7,0) -- node[h] {$j_{x,y-1,z}^{(2)}$} (-3.2,0);
\draw[l] (-4.7,-0.1) -- (-4.7,0.1);
\draw[l] (0.2,0) --  node[h] {$j_{x,y,z}^{(2)}$} (1.7,0);
\draw[l] (0.2,-0.1) -- (0.2,0.1);
\draw[l] (1.7,0) -- (2.3,0);
\draw[l] (2.45,0) circle(1.5mm); \node at (2.5,0.4) {$h_{^{x\!,y\!,z}}^{\!(2)}$};
\draw[l] (2.6,0) -- (3.2,0);
\draw[l] (3.2,0) -- node[h] {$j_{x,y,z}^{(2)}$} (4.7,0);
\draw[l] (4.7,-0.1) -- (4.7,0.1);
\draw[l] (0,0.9) --  node[v] {$j_{x,y,z}^{(3)}$} (0,1.7);
\draw[l] (-0.1,0.2) -- (0.1,0.2);
\draw[l] (0,0.2) -- (0,0.9);
\draw[l] (0,1.7) -- (0,2.3);
\draw[l] (0,2.45) circle(1.5mm); \node at (-0.5,2.4) {$h_{^{x\!,y\!,z}}^{\!(3)}$};
\draw[l] (0,2.6) -- (0,3.2);
\draw[l] (0,3.2) --  node[v] {$j_{x,y,z}^{(3)}$} (0,4.7);
\draw[l] (-0.1,4.7) -- (0.1,4.7);
\draw[l] (0,-1.7) --  node[v] {$j_{x,y,z-1}^{(3)}$} (0,-0.9);
\draw[l] (-0.1,-0.2) -- (0.1,-0.2);
\draw[l] (0,-0.9) -- (0,-0.2);
\draw[l] (0,-1.7) -- (0,-2.3);
\draw[l] (0,-2.45) circle(1.5mm); \node at (-0.55,-2.55) {$h_{^{x\!,y\!,z\texttt{-\!}1}}^{\!(3)}$};
\draw[l] (0,-2.6) -- (0,-3.2);
\draw[l] (0,-4.7) --  node[v] {$j_{x,y,z-1}^{(3)}$} (0,-3.2);
\draw[l] (-0.1,-4.7) -- (0.1,-4.7);
\node at (-2,2.5) {$\atopp{x\ direction}{\odot}$};
\node at (-2,1.5) {$\atopp{-x\ direction\ \ }{\otimes}$};
\end{tikzpicture}
},\!
\end{equation}
where the node $v_{x,y,z}$ is placed at $(x,y,z)$, and the symbol $j_{x+n,y,z}^{(i)}$ denotes the spin number attached to a link along the $i$-axis, which starts at the node $\{x\!+\!n,y,z\}$ and ends at the node $\{x\!+\!n\!+\!1,y,z\}$.  From now on, we assume the right-handed orientation of links, {\it i.e.} the link with the spin number $j_{x+n-1,y,z}^{(i)}$ is ingoing to the node $\{x+n,y,z\}$, while $j_{x+n,y,z}^{(i)}$ is outgoing from the same node.

In what follows, we will need the expression of the powers of the volume operator $\hat{\mathbf{V}}$, which acts diagonally \cite{Alesci:2013xd} on the reduced (and normalized) states \eqref{state_gravity}:
\begin{equation}
\begin{split}\label{volume}
\hat{\mathbf{V}}^n(v_{x,y,z})
\ket{\Gamma;j_l}_{\!R}
&=\left(8\pi\gamma l_P^2\right)^{\!\frac{3}{2}n}
\!\left(
\frac{j_{x-1,y,z}^{(1)}\!+j_{x,y,z}^{(1)}}{2}\ \frac{j_{x,y-1,z}^{(2)}\!+j_{x,y,z}^{(2)}}{2}\ \frac{j_{x,y,z-1}^{(3)}\!+j_{x,y,z}^{(3)}}{2}
\right)^{\!\!\!\frac{n}{2}}\!\!
\ket{\Gamma;j_l}_{\!R}
=\\
&=
\mathbf{V}^n_{\!v_{x\!,y\!,z}}
\ket{\Gamma;j_l}_{\!R},
\end{split}
\end{equation}
with $\gamma$ being the Immirzi parameter and
$\mathbf{V}_{v}:=\Big(\big(8\pi\gamma l_P^2\big)^{\!3}\,\Sigma_{v}^{(1)}\,\Sigma_{v}^{(2)}\,\Sigma_{v}^{(3)}\Big)^{\!\frac{1}{2}}$
being the eigenvalue of the volume operator $\hat{\mathbf{V}}$. The quantity
$\Sigma^{(i)}_{v}:=\frac{1}{2}\big(j^{(i)}_{v}+j^{(i)}_{v-\vec{e}_i}\big)$
denotes the averaged value of spins attached to the collinear pair of links (ingoing and outgoing) emanated from the node $v$, where $\vec{e}_{i}$ is the unit vector along the direction $i$, such that $j^{(i)}_{v-\vec{e}_i}$ represents the spin number of the link along $l^p$ (with $p$ being oriented along the fiducial direction $i$) ending in $v$. 


\section{Loop framework for quantum vector field}\label{III}
\noindent
The action of the gauge field $\underline{A}^I_\mu$ minimally coupled to gravity reads (we use the notation of \cite{Thiemann:1997rt})
\begin{equation}\label{action}
S^{(\underline{A})}=-\frac{1}{4Q^2}\int_M\!\!\!\!d^{4}x\sqrt{-g}g^{\mu\nu}g^{\rho\sigma}\underline{F}^I_{\mu\rho} \underline{F}^I_{\nu\sigma},
\end{equation}
where $Q^2$ is the coupling constant of dimension {\bf$1/\hbar$}, $g$ is the determinant of four-dimensional metric tensor and $\underline{F}^I_{\mu\rho}$ denotes the field strength 
\begin{equation}
\underline{F}^I_{\mu\nu}=\partial_\mu \underline{A}^I_{\nu}-\partial_\nu \underline{A}^I_{\mu} + C^{I}_{JK} \,\underline{A}^J_{\mu}\,\underline{A}^K_{\nu},
\end{equation}
$C^{I}_{JK}$ being structure constants of the gauge group.

The Legendre transform gives the following Hamiltonian:
\begin{equation}\label{Hamiltonian}
H^{(\underline{A})}\!=\!\int_{\Sigma_t}\!\!\!\!d^3x\bigg(\!
-\underline{A}^I_t\underline{D}_a\underline{E}^a_I
+N^a\underline{F}^I_{ab}\underline{E}^b_I
+N\frac{Q^2\!}{2\sqrt{q}}q_{ab}\left( \underline{E}^a_I\underline{E}^b_I+\underline{B}^a_I\underline{B}^b_I\right)
\!\bigg)
=\!\int_{\Sigma_t}\!\!\!\!d^3x\big(\underline{A}^I_t\underline{\mathcal{G}}_I^{(\underline{A})}
+N^a\mathcal{V}_a^{(\underline{A})}+N\mathcal{H}^{(\underline{A})}_{\text{sc}}\big),
\end{equation}
where $N$ is the lapse function and $N^a$ is the shift vector, while $\mathcal{V}_a^{(\underline{A})}$ and $\mathcal{H}^{(\underline{A})}_{\text{sc}}$ contribute to to the vector and scalar constraints densities of a gravitational field respectively. $\underline{\mathcal{G}}_I^{(\underline{A})}$ is the Yang-Mills field Gauss constraint dentity that generates gauge transformations in phase space. $\underline{E}_I^a=\frac{\sqrt{q}}{Q^2}e_0^{\mu}q^{ab}\underline{F}^J_{\mu b}\delta_{IJ}$ is the conjugate momentum to the vector field $\underline{A}^I_a$ and $\underline{B}^a_I=\frac{\sqrt{q}}{2Q^2}\epsilon^{abc}\underline{F}^J_{bc}\delta_{IJ}$ is its magnetic field, where $\epsilon^{abc}:=\frac{1}{\sqrt{q}}\tilde{\epsilon}^{abc}$ and $\tilde{\epsilon}^{abc}$ is the flat Levi-Civita symbol. We also introduced $q_{ab}$ denoting three-dimensional metric on the spacial Cauchy hypersurface $\Sigma_t$, while $q^{ab}$ and $q$ are the inverse and the determinant of the metric $q_{ab}$, respectively.

The vector field complement to the total vector constraint reads $\mathcal{V}_a^{(\underline{A})}:=\underline{F}^I_{ab}\underline{E}^b_I$. It generates diffeomorphism transformations.

The contribution to the smeared scalar constraint, $\mathcal{H}^{(\underline{A})}_{\text{sc}}[N]$ encodes all information about the dynamics of the vector field in the gauge and diffeomorphisms invariant phase-space and reads
\begin{equation}\label{Hamiltonianconstraint}
H^{(\underline{A})}_{\text{sc}}[N]:=\!\int_{\Sigma_t}\!\!\!\!d^3x
N\frac{Q^2\!}{2\sqrt{q}}q_{ab}\big(\underline{E}^a_I\underline{E}^b_I+\underline{B}^a_I\underline{B}^b_I\big)
:=H^{(\underline{A})}_{E}[N]+H^{(\underline{A})}_{B}[N],
\end{equation}
where we split it into two parts, the electric and magnetic ones, {\it i.e.}
\begin{align}
H^{(\underline{A})}_{E}[N]=&\ \frac{Q^2\!}{2}\!\int_{\Sigma_t}\!\!\!\!d^3x\frac{N}{\sqrt{q}}q_{ab}\,\underline{E}^a_I\underline{E}^b_I
\label{electric}
\\
H^{(\underline{A})}_{B}[N]=&\ \frac{Q^2\!}{2}\!\int_{\Sigma_t}\!\!\!\!d^3x\frac{N}{\sqrt{q}}q_{ab}\,\underline{B}^a_I\underline{B}^b_I
\label{magnetic}.
\end{align}

The phase space coordinates are given by holonomies of the gauge group along reduced graphs $\Gamma$, which we denote as follows 
\begin{equation}
\underline{h}_{\Gamma}\!:=\mathcal{P}\exp\!\left(\int_{\Gamma}\underline{A}_a\big(\Gamma(s)\big)\dot{\Gamma}^a(s)\right),
\end{equation}
where $\mathcal{P}$ denotes path ordering and $\underline{\tau}_I$ are the generators of the gauge group, and fluxes of the electric vector field around surfaces $S$, {\it i.e.}
\begin{equation}\label{smearedelectric}
\underline{E}_I(S^p):=\epsilon_{pqr}\!\int_{\!S\!\perp l^p\!(v)}\!\!\!\!\!\!\!\!\!\!\!\!dl^q\,dl^r\underline{E}_I^p(v),
\end{equation}
where we defined $S$ as the surface spanned by two lattice links $l^q$ and $l^r$, dual to $l^p$ (in the expression above only the indexes $q$ and $r$ are summed). In what follows we will only consider positive oriented links $l$, which provides positive oriented surfaces $S$ with respect to the fiducial directions.

In order to regularize the expression (\ref{Hamiltonianconstraint}) of the scalar constraint for the vector field (see section \ref{IV}), we need to express the electric and magnetic vector fields in terms of holonomies and fluxes. This can be done in the limit of finer and finer cubulation of the spatial manifold as follows. The electric field can be written in terms of fluxes using the following relation 
\begin{equation}\label{smearedelectric2}
\underline{E}_I(S^p)\approx\varepsilon^2\underline{E}_I^a(v)\,\delta_a^p\,,
\end{equation}
$\epsilon$ being the length of the links spanning $S^p$. As soon as the magnetic field $\underline{B}^a_I$ is concerned, one introduces the gauge holonomy
$\underline{h}_{q\circlearrowleft r}\big(\Delta(v)\big)
=\underline{h}_{l^r}^{-1}\big(\Delta(v)\big)\,\underline{h}_{l^q}^{-1}\big(\Delta(v)\big)
\,\underline{h}_{l^r}\big(\Delta(v)\big)\,\underline{h}_{l^q}\big(\Delta(v)\big)$
along a rectangular loop based at $v$ having links along the directions $q$ and $r$. In the limit of infinitesimal loops, whose links have length $\epsilon$, the holonomy $\underline{h}_{q\circlearrowleft r}$ can be expanded as follows
\begin{equation}\label{magncorrections}
\underline{h}_{q\circlearrowleft r}=1+\frac{1}{2}\varepsilon^2\underline{F}_{qr}+O(\varepsilon^4),
\end{equation}
where the curvature has been contracted with the gauge group generator, $\underline{F}_{qr}=\underline{F}_{qr}^I\tau_I$. In what follows, the following identity will be used
\begin{equation}\label{smearedmagnetic}
\frac{1}{Q^2}\epsilon^{pqr}\,\text{tr}\Big(\underline{\tau}_I\underline{h}_{q\circlearrowleft r}\big(\Delta(v)\big)\Big)
=\frac{\varepsilon^2}{2Q^2}\epsilon^{pqr}\,\text{tr}\big(\underline{\tau}_I\underline{F}_{qr}(v)\big)+O(\varepsilon^4)
\approx\frac{\varepsilon^2\underline{B}^a_I(v)}{\mathbf{V}(v,\varepsilon)}\,\delta_a^p\,,
\end{equation}
where $\underline{\tau}_I$ are the Yang-Mills generators in the fundamental representation, for which
\begin{equation}
\text{ tr}\big(\underline{\tau}_I\, \underline{\tau}_J\big)
=\delta_{IJ},
\end{equation}
and $\mathbf{V}(v,\varepsilon)$ is the volume of the region around $v$ with coordinate volume $\varepsilon^3$. { It is worth nothing that discretizing over the cuboidal lattice \eqref{state_gravity}, the continuous expression for the magnetic field
$\underline{B}^a_I=\frac{\sqrt{q}}{2Q^2}\epsilon^{abc}\underline{F}^J_{bc}\delta_{IJ}$
, one gets the sum of fields \eqref{smearedmagnetic} attached to the outgoing and ingoing links $l^p$ emanated from the node $v$, which can be reexpressed as the four gauge holonomies along loops dual to a given fiducial direction $i=p$ and based at the node $v$.}

We quantize the system of gravitational field and the vector field applying the method described in \cite{Thiemann:1997rt} for LQG to the case of QRLG. Therefore, the total Hilbert space is the direct product:
\begin{equation}
\mathcal{H}_{kin}^{(tot)}={}^{R\!}\mathcal{H}_{kin}^{(gr)}\!\otimes\mathcal{H}_{kin}^{(\underline{A})}\,,
\end{equation}
where the Hilbert space for gravity ${}^{R\!}\mathcal{H}_{kin}^{(gr)}$ has been described in the previous section. The Hilbert space $\mathcal{H}_{kin}^{(\underline{A})}$ for the gauge field is defined in terms of cylindrical functions of holonomies of the gauge connections, the only difference with \cite{Thiemann:1997rq} being that only reduced graphs are considered. Basis states of the full theory are represented as
$\ket{\Gamma;m_l;\underline{n}_l,\underline{i}_v}_{\!R}=\ket{\Gamma;m_l}_{\!R}\otimes\ket{\Gamma;\underline{n}_l,\underline{i}_v}$
, where $\ket{\Gamma;\underline{n}_l,\underline{i}_v}$ are invariant spin network states for the gauge field based at the graph $\Gamma$ and they are labeled by the quantum numbers $\underline{n}_l$ of the irreducible representation of the gauge group at each link $l$ and the corresponding invariant intertwiners $\underline{i}_v$ at nodes $v$.

The basic matter operators acting on $\ket{\Gamma;\underline{n}_l,\underline{i}_v}$ are the quantum gauge holonomies $\underline{\hat{h}}$, which act by multiplication in the same way as their gravitational equivalents, creating the holonomy of the composition of two paths, and the quantum electric fluxes that act as  the left/right invariant vector fields.

The states in the total Hilbert space $\mathcal{H}_{kin}^{(tot)}$ can be described in the following graphical way:
\begin{equation}\label{state}
\!\!\!\!\!\ket{\Gamma;j_l;\underline{n}_l,\underline{i}_v}_{\!R}=\!
\ket{\Gamma;j_l}_{\!R}\otimes\ket{\Gamma;\underline{n}_l,\underline{i}_v}
=\!\vket{\!
\begin{tikzpicture}
\draw[l] (-1.7,0.4) --  node[h] {$j_{x,y-1,z}^{(2)}$} (-0.9,0.4);
\draw[l] (-1.7,0.4) -- (-2.3,0.4);
\draw[l] (-4,0.3) -- (-4,0.5);
\draw[l] (-0.9,0.3) -- (-0.9,0.5);
\draw[l] (-2.45,0.4) circle(1.5mm); \node at (-2.4,0.8) {$h_{^{x\!,y\texttt{-\!}1\!,z}}^{\!(2)}$};
\draw[l] (-2.6,0.4) -- (-3.2,0.4);
\draw[l] (-4,0.4) -- node[h] {$j_{x,y-1,z}^{(2)}$} (-3.2,0.4);
\draw[l] (0.9,0.4) --  node[h] {$j_{x,y,z}^{(2)}$} (1.7,0.4);
\draw[l] (1.7,0.4) -- (2.3,0.4);
\draw[l] (4,0.3) -- (4,0.5);
\draw[l] (0.9,0.3) -- (0.9,0.5);
\draw[l] (2.45,0.4) circle(1.5mm); \node at (2.5,0.8) {$h_{^{x\!,y\!,z}}^{\!(2)}$};
\draw[l] (2.6,0.4) -- (3.2,0.4);
\draw[l] (3.2,0.4) -- node[h] {$j_{x,y,z}^{(2)}$} (4,0.4);
\draw[l] (-0.4,0.9) --  node[v] {$j_{x,y,z}^{(3)}$} (-0.4,1.7);
\draw[l] (-0.4,1.7) -- (-0.4,2.3);
\draw[l] (-0.5,0.9) -- (-0.3,0.9);
\draw[l] (-0.5,4) -- (-0.3,4);
\draw[l] (-0.4,2.45) circle(1.5mm); \node at (-0.9,2.4) {$h_{^{x\!,y\!,z}}^{\!(3)}$};
\draw[l] (-0.4,2.6) -- (-0.4,3.2);
\draw[l] (-0.4,3.2) --  node[v] {$j_{x,y,z}^{(3)}$} (-0.4,4);
\draw[l] (-0.4,-1.7) --  node[v] {$j_{x,y,z-1}^{(3)}$} (-0.4,-0.9);
\draw[l] (-0.4,-1.7) -- (-0.4,-2.3);
\draw[l] (-0.5,-0.9) -- (-0.3,-0.9);
\draw[l] (-0.5,-4) -- (-0.3,-4);
\draw[l] (-0.4,-2.45) circle(1.5mm); \node at (-0.95,-2.55) {$h_{^{x\!,y\!,z\texttt{-\!}1}}^{\!(3)}$};
\draw[l] (-0.4,-2.6) -- (-0.4,-3.2);
\draw[l] (-0.4,-4) --  node[v] {$j_{x,y,z-1}^{(3)}$} (-0.4,-3.2);
\draw[l] (-1.5,0) --  (0,0);
\node at (-0.75,-0.4) {$\underline{n}_{x,y-1,z}^{(2)}$};
\draw[l] (-1.5,0) -- (-1.7,0);
\draw[l] (-1.7,0) -- (-2.3,0);
\draw[l] (-2.45,0) circle(1.5mm);
\node at (-2.4,-0.45) {$\underline{h}_{^{x\!,y\texttt{-\!}1\!,z}}^{\!(2)}$};
\draw[l] (-2.6,0) -- (-3.2,0);
\draw[l] (-3.2,0) -- (-3.4,0);
\draw[l] (-4.9,0) -- (-3.4,0);
\node at (-4.15,-0.4) {$\underline{n}_{x,y-1,z}^{(2)}$};
\draw[l] (-4.9,-0.8) -- (-4.9,0.8);
\draw[l] (-4.9,0) -- (-5.1,0);
\draw[l] (0,0) -- (1.5,0);
\node at (0.75,-0.4) {$\underline{n}_{x,y,z}^{(2)}$};
\draw[l] (1.5,0) -- (1.7,0);
\draw[l] (1.7,0) -- (2.3,0);
\draw[l] (2.45,0) circle(1.5mm);
\node at (2.5,-0.45) {$\underline{h}_{^{x\!,y\!,z}}^{\!(2)}$};
\draw[l] (2.6,0) -- (3.2,0);
\draw[l] (3.2,0) -- (3.4,0);
\draw[l] (3.4,0) -- (4.9,0);
\node at (4.15,-0.4) {$\underline{n}_{x,y,z}^{(2)}$};
\draw[l] (4.9,-0.8) -- (4.9,0.8);
\draw[l] (4.9,0) -- (5.1,0);
\draw[l] (0,0.7) -- (0,1.5);
\node at (0.6,1.1) {$\underline{n}_{x,y,z}^{(3)}$};
\draw[l] (0,0) -- (0,0.7);
\draw[l] (0,1.5) -- (0,1.7);
\draw[l] (0,1.7) -- (0,2.3);
\draw[l] (0,2.45) circle(1.5mm);
\node at (0.6,2.4) {$\underline{h}_{^{x\!,y\!,z}}^{\!(3)}$};
\draw[l] (0,2.6) -- (0,3.2);
\draw[l] (0,3.2) -- (0,3.4);
\draw[l] (0,3.4) -- (0,4.9);
\node at (0.6,4.15) {$\underline{n}_{x,y,z}^{(3)}$};
\draw[l] (0,4.9) -- (0,5.1);
\draw[l] (-0.8,4.9) -- (0.8,4.9);
\draw[l] (0,-1.5) -- (0,-0.7);
\node at (0.8,-1.1) {$\underline{n}_{x,y,z-1}^{(3)}$};
\draw[l] (0,-0.7) -- (0,0);
\draw[l] (0,-1.5) -- (0,-1.7);
\draw[l] (0,-1.7) -- (0,-2.3);
\draw[l] (0,-2.45) circle(1.5mm);
\node at (0.65,-2.55) {$\underline{h}_{^{x\!,y\!,z\texttt{-\!}1}}^{\!(3)}$};
\draw[l] (0,-2.6) -- (0,-3.2);
\draw[l] (0,-3.2) -- (0,-3.4);
\draw[l] (0,-4.9) -- (0,-3.4);
\node at (0.8,-4.15) {$\underline{n}_{x,y,z-1}^{(3)}$};
\draw[l] (0,-5.1) -- (0,-4.9);
\draw[l] (0.8,-4.9) --  (-0.8,-4.9);
\node at (-2.5,3) {$\atopp{x\ direction}{\odot}$};
\node at (-2.5,2) {$\atopp{-x\ direction\ \ }{\otimes}$};
\end{tikzpicture}
},\!\!
\end{equation}
where $j_{p,q,r}^{(i)}$ and $n^{(i)}_{p,q,r}$ are the spin numbers at the associated links $l_{p,q,r}^{(i)}$ and $\underline{i}_v$ are the intertwiners at nodes.


\section{Regularization of scalar constraint}\label{IV}
\noindent
The quantization of the vector field part of the scalar constraint requires a regularization of the both terms in the formula \eqref{Hamiltonianconstraint}. This procedure is done still being in the classical phase space, by rewriting \eqref{Hamiltonianconstraint} in terms of holonomies and fluxes for both the gravitational part and the terms involving the vector field.

The gravitational part of Hamiltonian constraint is regularized by the method developed in LQG \cite{Thiemann:1996aw}, which has been restricted to cuboidal graphs \cite{Alesci:2013xd,Alesci:2015nja}.  Matter coupled to a dynamical spacetime is regularized introducing matter holonomies coupled to the dynamical lattice (composed of nodes and links). Alternatively, one can introduce a dual picture\cite{covariant}, where matter fields are coupled to the dynamics of granulated space (being volumes and areas of chunks of space) .

In order to regularize $H^{(\underline{A})}_E$ and $H^{(\underline{A})}_B$, we follow the procedure given in \cite{Thiemann:1997rt} for LQG in the presence of a vector field, making only restriction to the cuboidal graph.

At first, smearing electric vector fields and applying Thiemann's trick \cite{Thiemann:1996aw}, one has
\begin{equation}
e^i_a(x)=\frac{4}{n\gamma\kappa\big(\mathbf{V}(R)\big)^{\!n-1}}\Big\{A^i_a(x),\big(\mathbf{V}(R)\big)^{\!n}\Big\},
\end{equation}
such that one can write the following expression for the electric part of Hamiltonian
\begin{equation}
\begin{split}
H^{(\underline{A})}_{E}[N]
&=\frac{Q^2\!}{2}\lim_{\varepsilon\to0}\int\!\!d^3x\,N(x)
\frac{e^i_a(x)\underline{E}^a_I(x)}{\mathbf{V}^{\!\frac{1}{2}}(x,\varepsilon)}
\int\!\!d^3y\,
\frac{e^i_b(y)\underline{E}^b_I(y)}{\mathbf{V}^{\!\frac{1}{2}}(y,\varepsilon)}
\chi_{\varepsilon}(x,y)
=\\
&=
\frac{2^5Q^2}{(\gamma\kappa)^2}\lim_{\varepsilon\to0}\int\!\!d^3x\,N(x)
\big\{A^i_a(x),\mathbf{V}^{\frac{1}{2}}(x,\varepsilon)\big\}\underline{E}^a_I(x)
\!\int\!\!d^3y\,
\big\{A^i_b(y),\mathbf{V}^{\frac{1}{2}}(y,\varepsilon)\big\}\underline{E}^b_I(y)
\chi_{\varepsilon}(x,y)
\end{split}
\end{equation}
where it has been used the definition of the characteristic function $\chi_{\varepsilon}(x,y)$ of the box $B_{\varepsilon}(x)$ centered in $x$ with coordinate volume $\varepsilon^3$, precisely
\begin{equation}\label{volcorrections}
\mathbf{V}\big(B_{\varepsilon}(x)\big):=\mathbf{V}(x,\varepsilon)=\varepsilon^3\sqrt{q}(x)+O(\varepsilon^4),
\end{equation}
which allows to smear a function at the point $v$, around infinitesimal neighborhood, such that
\begin{equation}\label{characteristic}
f(x)=\int\!\!d^3y\,\delta^3(x-y)f(y)=\lim_{\varepsilon\to0}\frac{1}{\varepsilon^3}\!\int\!\!d^3y\,\chi_{\varepsilon}(x,y)f(y)\,.
\end{equation}

As a next step, we discretize the scalar constraint analogously to the case in which one uses a triangularization of the spatial manifold \cite{Thiemann:1996aw} for the gravitational field in presence of matter \cite{Thiemann:1997rt}. The idea that we follow is to replace the integration over the spatial hypersurface $\int_{\Sigma}$ with the sum over the ordered tetrahedra being the triples of links $\{l,l',l''\}$ emanating from all nodes $v$ of the graph $\Gamma$. In the case of a cubulation, each node $v$ is always surrounded by eight such triples and it is worth nothing that for each tetrahedron, the remaining seven ones coincide with the seven ``virtual'' tetrahedra, which are the objects that appear in the procedure of triangulation of any cuboidal or non-cuboidal lattice.

In other words, discretized integration over each tetrahedron, $\int_{\Delta_{l,l'\!,l''}}$, becomes the sum over the eight possibilities for choosing a triple of perpendicular links $\{l,l',l''\}$ among each tetrahedron of the triangulation $\Delta(v)$ around the node $v$. 

Using the triangulation method one gets the result 
\begin{equation}\label{regularizationE}
\begin{split}
H^{(\underline{A})}_{E}[N]
&=\frac{2^5Q^2}{(16\pi G\gamma)^2}\lim_{\varepsilon\to0}\varepsilon^6
\!\!\!\!\sum_{v\in\mathbf{V}(\Gamma)}\sum_{v'\in\mathbf{V}(\Gamma)\!}\,\sum_{\Delta(v)}\,\sum_{\Delta'(v')}\!N(v)
\big\{A^i_a(v),\mathbf{V}^{\frac{1}{2}}(v,\varepsilon)\big\}\delta^a_{l(v)}\,\underline{E}^c_I(v)\,\delta_c^{l(v)}\times\\
&\hspace{69.5mm}\times
\big\{A^i_b(v'),\mathbf{V}^{\frac{1}{2}}(v',\varepsilon)\big\}\delta^b_{l(v')}\,\underline{E}^d_I(v')\,\delta_d^{l(v')}
\delta_{v,v'}\approx\\
&\approx\frac{2^3Q^2}{(8\gamma\pi G)^2}\lim_{\varepsilon\to0}
\,\sum_{v,v'}\,\sum_{\Delta(v),\Delta'(v')\!}\!\!\!\!\!N(v)
\text{ tr}
\Big(
\tau^ih_{l^p}^{-1}\big(\Delta(v)\big)\big\{\mathbf{V}^{\frac{1}{2}}\big(\Delta(v)\big),\,h_{l^p}\big(\Delta(v)\big)\big\}
\!\Big)
\underline{E}_I\big(S^p(v)\big)\times\\
&\hspace{49mm}\times
\text{tr}
\Big(
\tau^ih_{l^s}^{-1}\big(\Delta'(v')\big)\big\{\mathbf{V}^{\frac{1}{2}}\big(\Delta'(v')\big),\,h_{l^s}\big(\Delta'(v')\big)\big\}
\!\Big)
\underline{E}_I\big(S^s(v')\big)\,
\delta_{v,v'}
\end{split}
\end{equation}
where $h_{l(\Delta)}$ are the SU(2) holonomies in the fundamental representation \cite{Thiemann:1996aw}, $\tau^j=-\frac{i}{2}\sigma^j$ are the algebra generators with $\sigma^j$ being Pauli matrices and tr denotes the trace over SU(2) algebra. The summations $\sum_{v\in\mathbf{V}(\Gamma)}$ and $\sum_{\Delta(v)}$ extend over all the nodes of the cubulation and over all the tetrahedra around each node, respectively, while the symbol $\delta^a_{l}$ means that $a$ is restricted to assume the value corresponding to the fiducial direction of the link $l$, {\it i.e.} for $l=l^p$ one has $a=p$. An additional factor $1/2^2$ has been added to account for the fact that each electric field has been smeared along both the positive and the negative oriented surfaces based at nodes, which are dual to the ingoing and outgoing links along the fiducial directions emanating from the nodes themselves. In the last line the following expansion
\begin{equation}\label{trick}
\,\text{tr}\big(\tau^ih_{l^a}^{-1}\big\{\mathbf{V}^{n}(R),h_{l^a}\big\}\big)
=-\,\text{tr}\big(\tau^i\varepsilon\big\{A_a,\mathbf{V}^{n}(R)\big\}+O(\varepsilon^2)\big)
\approx\frac{1}{2}\varepsilon\big\{A_a^i,\mathbf{V}^{n}(R)\big\}
\end{equation}
 has been applied and vector fields has been replaced by fluxes \eqref{smearedelectric}.

The magnetic part of Hamiltonian \eqref{magnetic} is regularized by the same method and reads:
\begin{equation}\label{smearedB}
H^{(\underline{A})}_{B}[N]=
\frac{2^5Q^2}{(\gamma\kappa)^2}\lim_{\varepsilon\to0}\int\!\!d^3x\,N(x)
\big\{A^i_a(x),\mathbf{V}^{\frac{1}{2}}(x,\varepsilon)\big\}\frac{\underline{B}^a_I(x)}{\mathbf{V}(x,\varepsilon)}
\!\int\!\!d^3y\,
\big\{A^i_b(y),\mathbf{V}^{\frac{1}{2}}(y,\varepsilon)\big\}\frac{\underline{B}^b_I(y)}{\mathbf{V}(y,\varepsilon)}
\chi_{\varepsilon}(x,y)\,.
\end{equation}
By using the same discretization adopted for the electric term \eqref{regularizationE}, one gets
\begin{equation}\label{regularizationB}
\begin{split}
H^{(\underline{A})}_{B}[N]
&=\frac{2^5Q^2}{(16\pi G\gamma)^2}\lim_{\varepsilon\to0}\varepsilon^6
\!\!\!\!\sum_{v\in\mathbf{V}(\Gamma)}\sum_{v'\in\mathbf{V}(\Gamma)\!}\,\sum_{\Delta(v)}\,\sum_{\Delta'(v')}\!N(v)
\big\{A^i_a(v),\mathbf{V}^{\frac{1}{2}}(v,\varepsilon)\big\}\delta^a_{l(v)}\frac{\underline{B}^c_I(v)}{\mathbf{V}(v,\varepsilon)}\,\delta_c^{l(v)}\times\\
&\times
\big\{A^i_b(v'),\mathbf{V}^{\frac{1}{2}}(v',\varepsilon)\big\}\delta^b_{l(v')}\frac{\underline{B}^d_I(v')}{\mathbf{V}(v',\varepsilon)}\,\delta_d^{l(v')}
\delta_{v,v'}\approx\\
&\approx\frac{2^5}{Q^2(8\gamma\pi G)^2}\lim_{\varepsilon\to0}
\,\sum_{v,v'}\,\sum_{\Delta(v),\Delta'(v')\!}\!\!\!\epsilon^{pqr}\epsilon^{stu}N(v)
\times\\&\times
\,\text{tr}
\Big(
\tau^ih_{l^p}^{-1}\big(\Delta(v)\big)\big\{\mathbf{V}^{\frac{1}{2}}\big(\Delta(v)\big),\,h_{l^p}\big(\Delta(v)\big)\big\}
\!\Big)
\,\text{tr}\Big(\underline{\tau}_I\underline{h}_{q\circlearrowleft r}\big(\Delta(v)\big)\Big)
\times\\&\times
\,\text{tr}
\Big(
\tau^ih_{l^s}^{-1}\big(\Delta'(v')\big)\big\{\mathbf{V}^{\frac{1}{2}}\big(\Delta'(v')\big),\,h_{l^s}\big(\Delta'(v')\big)\big\}
\!\Big)
\,\text{tr}\Big(\underline{\tau}_I\underline{h}_{q\circlearrowleft r}\big(\Delta'(v')\big)\Big)\,
\delta_{v,v'}
\end{split}
\end{equation}
where magnetic fields has been replaced by traces over generators and gauge holonomies \eqref{smearedmagnetic} of the Yang-Mills gauge group. 


\section{Quantization of the scalar constraint}\label{V}
\noindent
The canonical procedure of quantization of the field contribution to the scalar constraint, after the cubulation of the spatial manifold (which becomes the graph $\Gamma$ at which the state is based: links and nodes of the cubulations turns to links and nodes of $\Gamma$), is nothing but a change of holonomies, volumes and matter variables into quantum operators that act on states \eqref{state} belonging to $\mathcal{H}_{kin}^{(tot)}$:
\begin{equation}\label{HCO1}
\hat{H}^{(\underline{A})}\!\ket{\Gamma;j_l;\underline{n}_l,\underline{i}_v}_{\!R}
=\Big(\hat{H}^{(\underline{A})}_{E}+\hat{H}^{(\underline{A})}_{B}\Big)\!\ket{\Gamma;j_l;\underline{n}_l,\underline{i}_v}_{\!R}.
\end{equation}
The Poisson brackets in \eqref{regularizationE} and \eqref{regularizationB} are replaced by commutators times $\frac{1}{i\hbar}$ that appear only as elements of the following term:
\begin{equation}
\text{tr}
\Big(
\tau^i\hat{h}_{l^p(\Delta)}^{-1}\big[\hat{\mathbf{V}}^n(\Delta),\,\hat{h}_{l^p(\Delta)}\big]
\!\Big)=\,
\text{tr}
\Big(
\tau^i\hat{h}_{l^p(\Delta)}^{-1}\hat{\mathbf{V}}^n(\Delta)\,\hat{h}_{l^p(\Delta)}
\!\Big)\,.
\end{equation}

The quantum operator corresponding to the electric part \eqref{regularizationE} acts as follows:
\begin{equation}\label{operatorE1}
\begin{split}
\hat{H}^{(\underline{A})}_{E}\!\ket{\Gamma;j_l;\underline{n}_l,\underline{i}_v}_{\!R}
&=-\frac{2^3Q^2}{(8\gamma\pi l_P^2)^2}\lim_{\varepsilon\to0}
\,\sum_{v}\sum_{\Delta(v),\Delta'(v)\!\!}\!\!\!\!N(v)
\text{ tr}
\Big(
\tau^i\hat{h}_{l^p}^{-1}\big(\Delta(v)\big)\hat{\mathbf{V}}^{\frac{1}{2}}\big(\Delta(v)\big)\,\hat{h}_{l^p}\big(\Delta(v)\big)
\!\Big)
\underline{\hat{E}}_I\big(S^p(v)\big)\times\\
&\times
\,\text{tr}
\Big(
\tau^i\hat{h}_{l^s}^{-1}\big(\Delta'(v)\big)\hat{\mathbf{V}}^{\frac{1}{2}}\big(\Delta'(v)\big)\,\hat{h}_{l^s}\big(\Delta'(v)\big)
\!\Big)
\underline{\hat{E}}_I\big(S^s(v)\big)\!\ket{\Gamma;j_l;\underline{n}_l,\underline{i}_v}_{\!R}\,
\end{split}
\end{equation}
where we applied the Kronecker delta ($v=v'$) and we introduced the symbol $l_P=\sqrt{\hbar G}$ to denote Planck length. Note that the sum over triangulations ${\Delta(v),\Delta'(v)}$ does not depend on the choice of internal directions of gravitational su(2) group ($i$-indexes) and vector field gauge group ($I$-indexes), therefore it extends only over possible selections of tetrahedra which contain the link $l^p\big(\Delta(v)\big)$ and $l^s\big(\Delta'(v)\big)$ for every choice of direction $p$ and $s$ respectively.

It is worth nothing that in the limit $\varepsilon\to0$ the expression \eqref{operatorE1} gives finite outcome, hence the dependency on the regulator $\varepsilon$ can be simply removed. As a result one obtains a sum of subsystems, called basic cells\footnote{The example, with the central node $v_{x,y,z}$, is given by the illustration of state \eqref{state}.} extending over all nodes of the graph $\Gamma$. Each basic cell, labeled by the position of the central node, is a sum of elements acting on a it and the surrounding six nearest neighbor nodes, placed at the endpoints of links. This cellular structure allows to restrict calculations to a basic cell and to give the final result as the sum over all cells.

One can calculate the action of the operator \eqref{operatorE1} from the following expression (see appendix \eqref{A}):
\begin{equation}\label{traceaction}
\text{tr}\big(\tau^i\hat{h}_{l^p}^{-1}\hat{\mathbf{V}}^n(v)\,\hat{h}_{l^p}\big)\!\ket{\Gamma;j_l;\underline{n}_l,\underline{i}_v}_{\!R}
=\frac{i}{2}\left(8\pi\gamma l_P^2\right)^{\!\frac{3}{2}n}
\Delta^{(p),\frac{n}{2}}_{v}\,\delta^{i,p}\Big(\Sigma^{(j)}_{v}\Sigma^{(k)}_{v}\Big)^{\!\frac{n}{2}}
\ket{\Gamma;j_l;\underline{n}_l,\underline{i}_v}_{\!R},
\end{equation}
where $j,k$ denotes the two orthogonal directions to $i$ and \footnote{We are assuming all spin numbers $j^{(i)}$ are bigger than $1/2$, otherwise we get different expressions of $\Delta^{(p)}_v\delta^{i,p}$ for ingoing and outgoing links $l^p$.}   
\begin{equation}\label{Deltaexpansion}
\Delta^{(p),n}_v\,\delta^{i,p}=
\frac{1}{2^n}\bigg(\!\Big(j^{(i)}_v+j^{(i)}_{v-\vec{e}_i}-\frac{1}{2}\Big)^{\!n}
\!-\Big(j^{(i)}_v+j^{(i)}_{v-\vec{e}_i}+\frac{1}{2}\Big)^{\!n}\bigg)\delta^{i,p}
\end{equation}
Hence, the operator \eqref{operatorE1} reads:
\begin{equation}\label{operatorE2}
\begin{split}
\hat{H}^{(\underline{A})}_{E}\!\ket{\Gamma;j_l;\underline{n}_l,\underline{i}_v}_{\!R}
&=\frac{2Q^2}{(8\gamma\pi l_P^2)^{\frac{1}{2}}}\sum_{v}N(v)\,\delta_{p,s}
\sum_{T(v)}\,\sum_{[l^p,l^q,l^r]\in T(v)}\!\!\!\!
\Big(\Sigma^{(p)}_{v}\Sigma^{(q)}_{v}\Sigma^{(r)}_{v}\Big)^{\!\frac{1}{4}}
\frac{\Delta^{(p),\frac{1}{4}}_{v}}{\big(\Sigma^{(p)}_{v}\big)^{\!\frac{1}{4}}}
\underline{\hat{E}}_I\big(S^p(v)\big)
\times\\
&\times
\!\sum_{T'(v)}\,\sum_{[l^s,l^t,l^u]\in T'(v)}\!\!\!\!
\Big(\Sigma^{(s)}_{v}\Sigma^{(t)}_{v}\Sigma^{(u)}_{v}\Big)^{\!\frac{1}{4}}
\frac{\Delta^{(s),\frac{1}{4}}_{v}}{\big(\Sigma^{(s)}_{v}\big)^{\!\frac{1}{4}}}
\underline{\hat{E}}_I\big(S^s(v)\big)
\!\ket{\Gamma;j_l;\underline{n}_l,\underline{i}_v}_{\!R}.
\end{split}
\end{equation}

The summation $\sum_{[l^p,l^q,l^r]}$ extends over the 6 permutations of links in a given triple of edges, while the summations $\sum_{T(v)}$ goes over the 8 possible choices of triples with mutual orthogonal links. Note that the expression
$\big(\Sigma^{(p)}_{v}\Sigma^{(q)}_{v}\Sigma^{(r)}_{v}\big)^{\!\frac{1}{4}}$
is invariant under the choice of triple of links, hence one can put it in front of the sum as
$\big(\Sigma^{(1)}_{v}\Sigma^{(2)}_{v}\Sigma^{(3)}_{v}\big)^{\!\frac{1}{4}}$.
Each term of the summations above depend just on the internal direction $p$, such that the summation over the 6 permutations becomes the summation over the three fiducial directions times 2, $\sum_{[l^p]\in T(v)}\to2\sum_{i=1}^3$ and it gives the same result for each of the 8 choices of triples of links, hence $\sum_{T(v)}\to8$.

Finally, we end up with:
\begin{equation}\label{operatorE3}
\begin{split}
\!\!\hat{H}^{(\underline{A})}_{E}\!\ket{\Gamma;j_l;\underline{n}_l,\underline{i}_v}_{\!R}
&=
2^5\,Q^2(8\gamma\pi l_P^2)
\sum_{v}N(v)\frac{\Sigma^{(1)}_{v}\Sigma^{(2)}_{v}\Sigma^{(3)}_{v}}{\mathbf{V}_{v}}\sum_{i=1}^3\!
\Bigg(\frac{\Delta^{(i),\frac{1}{4}}_{v}}{\big(\Sigma^{(i)}_{v}\big)^{\!\frac{1}{4}}}\Bigg)^{\!\!2}
\times\\
&\times
\Big[\underline{\hat{E}}_I\big(S^i(v)\big)\Big]^2
\ket{\Gamma;j_l;\underline{n}_l,\underline{i}_v}_{\!R}
\,.
\end{split}
\end{equation}

Similarly, the quantization of the magnetic part of the scalar constraint (\ref{regularizationB}) results with the expression:
\begin{equation}\label{operatorB1}
\begin{split}
\hat{H}^{(\underline{A})}_{B}\!\ket{\Gamma;j_l;\underline{n}_l,\underline{i}_v}_{\!R}
&=-\frac{2^5}{Q^2(8\gamma\pi l_P^2)^2}\lim_{\varepsilon\to0}
\,\sum_{v}\sum_{\Delta(v),\Delta'(v)\!\!}\!\!\!\!\epsilon^{pqr}\epsilon^{stu}N(v)
\times\\
&\times
\,\text{tr}
\Big(
\tau^i\hat{h}_{l^p}^{-1}\big(\Delta(v)\big)\big[\hat{\mathbf{V}}^{\frac{1}{2}}\big(\Delta(v)\big),\,\hat{h}_{l^p}\big(\Delta(v)\big)\big]
\!\Big)
\,\text{tr}\Big(\underline{\tau}_I\underline{\hat{h}}_{q\circlearrowleft r}\big(\Delta(v)\big)\Big)
\times\\
&\times
\,\text{tr}
\Big(
\tau^i\hat{h}_{l^s}^{-1}\big(\Delta'(v)\big)\big[\hat{\mathbf{V}}^{\frac{1}{2}}\big(\Delta'(v)\big),\,\hat{h}_{l^s}\big(\Delta'(v)\big)\big]
\!\Big)
\,\text{tr}\Big(\underline{\tau}_I\underline{\hat{h}}_{t\circlearrowleft u}\big(\Delta'(v)\big)\Big)
\!\ket{\Gamma;j_l;\underline{n}_l,\underline{i}_v}_{\!R}
=\\
&=\frac{2^3}{Q^2(8\gamma\pi l_P^2)^{\frac{1}{2}}}
\,\sum_{v}N(v)\,\delta_{p,s}
\times\\
&\times
\sum_{T(v)}\,\sum_{[l^p,l^q,l^r]\in T(v)}\!\!\!\!\epsilon^{pqr}
\Big(\Sigma^{(p)}_{v}(\Sigma^{(q)}_{v}\Sigma^{(r)}_{v}\Big)^{\!\frac{1}{4}}
\frac{\Delta^{(p),\frac{1}{4}}_{v}}{\big(\Sigma^{(p)}_{v}\big)^{\!\frac{1}{4}}}
\,\text{tr}\Big(\underline{\tau}_I\underline{\hat{h}}_{q\circlearrowleft r}\big(T(v)\big)\Big)
\times\\
&\times
\sum_{T'(v)}\,\sum_{[l^s,l^t,l^u]\in T'(v)}\!\!\!\!\epsilon^{stu}
\Big(\Sigma^{(s)}_{v}\Sigma^{(t)}_{v}\Sigma^{(u)}_{v}\Big)^{\!\frac{1}{4}}
\frac{\Delta^{(s),\frac{1}{4}}_{v}}{\big(\Sigma^{(s)}_{v}\big)^{\!\frac{1}{4}}}
\,\text{tr}\Big(\underline{\tau}_I\underline{\hat{h}}_{t\circlearrowleft u}\big(T'(v)\big)\Big)
\!\ket{\Gamma;j_l;\underline{n}_l,\underline{i}_v}_{\!R},
\end{split}
\end{equation}
where we introduced the gauge holonomy operator operator, $\underline{\hat{h}}_{q\circlearrowleft r}\big(\Delta(v)\big)$ that is the quantum equivalent of the holonomy $\underline{h}_{q\circlearrowleft r}\big(\Delta(v)\big)$ \eqref{smearedmagnetic}.

The action of the operator \eqref{operatorB1} is derived as for the electric part of the scalar constraint operator \eqref{operatorE2}. The only nontrivial difference comes from the term
$\epsilon^{pqr}\epsilon^{ptu}\,\text{tr}\big(\underline{\tau}_I\underline{\hat{h}}_{q\circlearrowleft r}(T)\big)
\,\text{tr}\big(\underline{\tau}_I\underline{\hat{h}}_{t\circlearrowleft u}(T')\big)$,
which depends on the considered links for each fiducial direction $p\to i$. This operator at a given node is the summation over all the links emanating from the node of some terms which provide the insertion of the holonomies along two of the dual loops to the considered link. {Since there are two possibilities for choosing an orientation of the link $l^p$ along a given direction $i$ and the term
$\Delta^{(p),\frac{1}{4}}_{v}/\big(\Sigma^{(p)}_{v}\big)^{\!\frac{1}{4}}$
is symmetric under change of an orientation, hence one can simplify the summation
$\sum_{T(v)}\sum_{[l^p,l^q,l^r]\in T(v)}$ into $2\sum_{\{l^j,l^k\}\bot i}\sum_{i=1}^3$, where the summation $\sum_{\{l^j,l^k\} \bot i}$ extends over all orthogonal links to the fiducial direction $i$.} Therefore we end up with the action of the operator \eqref{operatorB1} expressed by the following formula:
\begin{equation}\label{operatorB2}
\begin{split}
\hat{H}^{(\underline{A})}_{B}\!\ket{\Gamma;j_l;\underline{n}_l,\underline{i}_v}_{\!R}
&=\frac{2^5}{Q^2}(8\gamma\pi l_P^2)
\sum_{v}N(v)\,\frac{\Sigma^{(1)}_{v}\Sigma^{(2)}_{v}\Sigma^{(3)}_{v}}{\mathbf{V}_{v}}
\sum_{i=1}^3
\!\Bigg(\frac{\Delta^{(i),\frac{1}{4}}_{v}}{\big(\Sigma^{(i)}_{v}\big)^{\!\frac{1}{4}}}\Bigg)^{\!\!2}
\times\\
&\times
\!\!\!\!\sum_{\{l^j,l^k\}\bot i}\,\sum_{\{l^l,l^m\}\bot i}\!\!\!\!\epsilon^{ijk}\epsilon^{ilm}
\,\text{tr}\big(\underline{\tau}_I\underline{\hat{h}}_{l^j\circlearrowleft l^k}(v)\big)\,
\,\text{tr}\big(\underline{\tau}_I\underline{\hat{h}}_{l^l\circlearrowleft l^m}(v)\big)
\!\ket{\Gamma;j_l;\underline{n}_l,\underline{i}_v}_{\!R},
\end{split}
\end{equation} 
where $\underline{\hat{h}}_{l^j\circlearrowleft l^k}$ denotes the holonomy along the square constructed from $l^j$ and $l^k$.


\subsection{Large \textit{j} limit}
\noindent
We now show how to perform the large $j$ limit of the formulas \eqref{operatorE3}, \eqref{operatorB2} and we outline how the expectation value of the quantum Hamiltonian coincides with the classical expression \eqref{Hamiltonianconstraint} at the leading order, as soon as a suitable semiclassical limit is performed. To calculate this limit one can use the definition of the characteristic function \eqref{characteristic} to smear discrete expressions and consider the following expansion for $j\gg\frac{1}{2}$\footnote{It is worth noting that one can remove the positive spin numbers restriction. Therefore, assuming $|j|>1/2$ in \eqref{Deltaexpansion}, one gets the same expression \eqref{gravcorrections} for $|j|\gg\frac{1}{2}$.}:
\begin{equation}\label{gravcorrections}
\Delta_{v}^{\!(p),n}=-\frac{n}{2}\big(\Sigma_{v}^{(p)}\big)^{n-1}\!+O\big(j^{n-3}\big)
\approx-\frac{n}{2}\bigg(\frac{p^{i}(v)\,\varepsilon^2}{8\pi\gamma l_P^2}\bigg)^{\!n-1},
\end{equation}
where $p^{(i)}(u)$ denote gravitational momenta at the point $u$, which are related to spin-numbers by the following relation  
\begin{equation}\label{Sigma_relation}
p^{i}(v)\,\varepsilon^2\!=8\pi\gamma l_P^2\Sigma^{(i)}_{v}.
\end{equation}

It is worth nothing that at the leading order the expansion \eqref{gravcorrections} gives terms of order $\Big(\frac{8\gamma\pi l_P^2}{\varepsilon^2p^i}\Big)^{\!2}$ and similar terms, namely of $\varepsilon^{-4}$ order, come form the smeared magnetic operators \eqref{magncorrections}. Any other artifacts of discretization (\textit{e.g.} the ones from the formulas \eqref{volcorrections} or \eqref{trick}) would give negligible contributions, providing additional positive $\varepsilon$ powers in the numerator.

Then the expectation value, 
${}_{\raisebox{-2.5pt}{\scriptsize$R$\!}}\big<\hat{H}^{(\underline{A})}\big>_{\!\!R}:=
{}_{\raisebox{-1.5pt}{\scriptsize$R\!\!$}}\bra{\Gamma;j_l;\underline{n}_l,\underline{i}_v}\hat{H}^{(\underline{A})}\!\ket{\Gamma;j_l;\underline{n}_l,\underline{i}_v}_{\!R}$
reads:
\begin{equation}
\begin{split}
{}_{\raisebox{-2.5pt}{\scriptsize$R$\!}}\big<\hat{H}^{(\underline{A})}\big>_{\!\!R}
\approx&\ 
\frac{Q^2}{2}
\lim_{\varepsilon\to0}\sum_{v}\frac{1}{\varepsilon^3}\!\int\!\!d^3u\,\chi_{\varepsilon}(v,u)\,\frac{N(v)}{\varepsilon^3\sqrt{q(v)}}
\sum_{i=1}^3
\frac{\varepsilon^2\,p^{1}(v)\,p^{2}(v)\,p^{3}(v)}{\big(p^{i}(v)\big)^{\!2}}
\times\\
&\times
\bigg(
\Big<\big[\underline{\hat{E}}_I\big(S^i(v)\big)\big]^2\Big>
+
\frac{1}{Q^4}\,q(v)\!\!\!\!\sum_{\{l^j,l^k\}\bot i}\,\sum_{\{l^l,l^m\}\bot i}\!\!\!\!\epsilon^{ijk}\epsilon^{ilm}
\Big<
\,\text{tr}\big(\underline{\tau}_I\underline{\hat{h}}_{l^j\circlearrowleft l^k}(v)\big)\,
\,\text{tr}\big(\underline{\tau}_I\underline{\hat{h}}_{l^l\circlearrowleft l^m}(v)\big)
\Big>
\bigg),
\end{split}
\end{equation}
where $q_{11}=\frac{p^{2}p^{3}}{p^{1}},\ q_{22}=\frac{p^{3}p^{1}}{p^{2}},\ q_{33}=\frac{p^{1}p^{2}}{p^{3}}$
are metric components, while $q=|p^{1}p^{2}p^{3}|$ is the determinant and each holonomy acts as a left or right invariant vector field.

Let us assume to construct a proper semiclassical state for the gauge field variables, such that expectation values and eigenvalues become classical quantities. Hence we get:
\begin{equation}
\begin{split}\label{limit}
\Big<\big[\underline{\hat{E}}_I\big(S^i(v)\big)\big]^2\Big>
&\leadsto\big[\underline{E}_I\big(S^i(v)\big)\big]^2
\\
\sum_{\{l^j,l^k\}\bot i}\,\sum_{\{l^l,l^m\} \bot i}\!\!\!\!\epsilon^{ijk}\epsilon^{ilm}
\Big<
\,\text{tr}\big(\underline{\tau}_I\underline{\hat{h}}_{l^j\circlearrowleft l^k}(v)\big)\,
\,\text{tr}\big(\underline{\tau}_I\underline{\hat{h}}_{l^l\circlearrowleft l^m}(v)\big)
\Big>
&\leadsto
\frac{\varepsilon^2}{2}
\,\text{tr}\big(\underline{\tau}_I\underline{F}_{jk}(v)\big)
\frac{\varepsilon^2}{2}
\,\text{tr}\big(\underline{\tau}_I\underline{F}_{lm}(v)\big),
\end{split}
\end{equation}
Note that the right hand sides of the expressions \eqref{limit} contribute to the smeared classical objects inside the box centered at $v$.

Then in the limit $\varepsilon\rightarrow 0$ we have $v=u$ and $\sum_{v}\!\int\!\!d^3u\,\chi_{\varepsilon}(v,u)=\int\!\!d^3u$, so finding
\begin{equation}\label{semiclassical}
\begin{split}
{}_{\raisebox{-2.5pt}{\scriptsize$R$\!}}\big<\hat{H}^{(\underline{A})}\big>_{\!\!R}
\leadsto
h^{(\underline{A})}
\approx&\ 
\frac{Q^2}{2}
\!\int\!\!d^3u\,\frac{N(u)}{\sqrt{q(u)}}
\sum_{i=1}^3
q_{ii}(u)
\bigg(
\big(\underline{E}_I^i(u)\big)^{\!2}
+
\epsilon^{ijk}\frac{\sqrt{q(u)}}{2Q^2}\underline{F}_{jk}^I(u)\,\epsilon^{ilm}\frac{\sqrt{q(u)}}{2Q^2}\underline{F}_{lm}^I(u)
\bigg),
\end{split}
\end{equation}
where the discrete eigenvalues has been replaced by the continuous variables. It precisely coincides with the classical expression \eqref{Hamiltonianconstraint} with the metric in the diagonal gauge.

\section{Conclusions}\label{IX}
\noindent
We extended the formulation of QRLG in order to include a gauge vector field. We settled down all the necessary tools in order to have a well-defined quantum theory, which essentially reduces to a lattice gauge theory on a cubic lattice. The adherence to the loop quantization program implied a peculiar expression for the matter part of the scalar constraint operator, which has been defined and analyzed, showing how it provides the right semiclassical limit as soon as proper semiclassical states for the gauge field are provided and a large-$j$ limit is taken for the gravitational degrees of freedom. 
 
Next-to-the-leading order terms in the large-$j$ expansion can be easily computed starting from the achievements of the present work and they provide the first kind of quantum gravity corrections computed for a vector field in LQG. This will be done in future developments. 

However, there are other kind of quantum gravity corrections, coming directly from the fact that the quantization of vector fields that has been implemented is not equivalent to the Fock quantization. The determination of these corrections would give us a comprehensive description of quantum vector fields on a quantum spacetime. 

The present analysis provides the expression of the quantum operator associated with the matter part of the scalar constraint, which generates the dynamics of the vector field on a quantum spacetime. The investigation of such dynamics is affected by the same kind of problems which plague the formulation of lattice gauge theory, namely the lack of explicit solutions, unless in some quite trivial cases. 

In this respect, the combination of the present results with the definition of a dynamical vacuum out of the Fock vacuum, given in \cite{Ashtekar:2001xp}, is a promising perspective in view of the application of the present framework to physically relevant cases.


\appendix
\section{}\label{A}
\noindent 
Below we refine the derivation provided in \cite{Bilski:2015dra} of the action of $\text{tr}\big(\tau^i\hat{h}_{l^p}^{-1}\hat{\mathbf{V}}^n(v)\,\hat{h}_{l^p}\big)$, so to recover \eqref{traceaction}. Let us perform the calculation for an outgoing link $l^p$ and $p=3$:
\begin{equation}
\text{tr}\big(\tau^i\hat{h}_{l^3}^{-1}\hat{\mathbf{V}}^n(v)\,\hat{h}_{l^3}\big)=-\sum_{abd}\,(\tau_i)_{ab}\, (\hat{h}^{-1}_{l^3})_{bd}\,V^n\,(\hat{h}_{l^3})_{da}
\end{equation}
where $a,b,d$ are indexes in the fundamental representation; let us choose the basis in which $\tau^3$ is diagonal and the holonomies read
$(\hat{h}_{l^3})_{da}=e^{ia\theta} \delta_{da}$.

Since the volume acts after the insertion of the holonomy $\hat{h}_{l^3}$, the application of the aforementioned operator to a state provides the coefficient
$[\Sigma^{(1)}\,\Sigma^{(2)}\,(\Sigma^{(3)}+a/2)]^{n/2}$.
Thus we find
\begin{equation}
=-\big(\Sigma^{(1)}\,\Sigma^{(2)}\big)^{\!\frac{n}{2}}\!\!\!\!\!\!\sum_{abd=-1/2}^{1/2}\!\!\!\!
(\tau_i)_{ab}\, (\hat{h}^{-1}_{l^3})_{bd}
\,\Big(\Sigma^{(3)}+\frac{a}{2}\Big)^{\!\frac{n}{2}}\,(\hat{h}_{l^3})_{da}
=
-\big(\Sigma^{(1)}\,\Sigma^{(2)}\big)^{\!\frac{n}{2}}\!\!\!\!\!\!\sum_{abd=-1/2}^{1/2}\!\!\!\!
(\tau_i)_{ab}\, e^{-id\theta} \delta_{bd}
\,\Big(\Sigma^{(3)}+\frac{a}{2}\Big)^{\!\frac{n}{2}}e^{ia\theta}\delta_{da}.
\end{equation}
Hence, using the $\delta$'s it turns out that $a=d=b$, such that the two exponentials disappear and
\begin{equation}
=-\big(\Sigma^{(1)}\,\Sigma^{(2)}\big)^{\!\frac{n}{2}}\sum_a
(\tau_i)_{aa}\,\Big(\Sigma^{(3)}+\frac{a}{2}\Big)^{\!\frac{n}{2}}.
\end{equation}
Since, the only generator with nonvanishing diagonal components is $\tau^3$, the expression above recasts
\begin{equation}
=-i\big(\Sigma^{(1)}\,\Sigma^{(2)}\big)^{\!\frac{n}{2}}\,\delta^{i,3} \sum_a
a\,\Big(\Sigma^{(3)}+\frac{a}{2}\Big)^{\!\frac{n}{2}}
=
\frac{i}{2}\big(\Sigma^{(1)}\,\Sigma^{(2)}\big)^{\!\frac{n}{2}}\,\delta^{i,3}
\Bigg[\bigg(\Sigma^{(3)}-\frac{1}{4}\bigg)^{\!\!\frac{n}{2}}-\bigg(\Sigma^{(3)}+\frac{1}{4}\bigg)^{\!\!\frac{n}{2}}\Bigg]\,,
\end{equation}
from which for outgoing link in the $p=3$ direction \eqref{traceaction} follows. 
For $p=1,2$, $h_{l^p}$ is diagonal modulo some discrete rotations, which can be moved to $\tau^i$. Therefore, the same result is obtained by  rotating the SU$(2)$ generator $\tau_i$ and the only non-vanishing contributions are for $i=1,2$. For ingoing link the only difference is that $(\hat{h}_{l^3})_{da}=e^{-ia\theta} \delta_{da}$ and the rest of the analysis is similar.



{\acknowledgments
J.~B. and A.~M. wishes to acknowledge support by the Shanghai Municipality, through the grant No. KBH1512299, and by Fudan University, through the grant No. JJH1512105. E.~A. wishes to acknowledge the John Templeton Foundation for the supporting grant \#51876. The work of F.~C. was supported by funds provided by the National Science Center under the agreement DEC12 2011/02/A/ST2/00294.}

\end{document}